\documentclass[aps,prx,twocolumn,superscriptaddress, nofootinbib,  longbibliography]{revtex4-2} 
\usepackage{graphicx} 
\usepackage{dcolumn}  
\usepackage{bm}    
\usepackage{amssymb}  
\usepackage{amsfonts, amsmath, amssymb, mathtools}
\usepackage{amsmath}

\usepackage{lipsum}
\usepackage[normalem]{ulem}
\usepackage{commath}
\usepackage{xcolor}
\usepackage{braket}

\begin{document}

\raggedbottom

\title{Realization of versatile and effective quantum metrology using a single bosonic mode}

\author{Xiaozhou Pan}
\thanks{Contributed equally to this article.\\
Corresponding author: xiaozhou@nus.edu.sg}
\author{Tanjung Krisnanda }
\thanks{Contributed equally to this article.\\
Corresponding author:
tanjung@nus.edu.sg}
\author{Andrea Duina}
\affiliation{Centre for Quantum Technologies, National University of Singapore, Singapore 117543, Singapore}
\author{Kimin Park}
\affiliation{Department of Optics, Palack\'{y} University, 17. listopadu 1192/12, 77146 Olomouc, Czech Republic}
\author{Pengtao Song}
\affiliation{Centre for Quantum Technologies, National University of Singapore, Singapore 117543, Singapore}
\author{Clara Yun Fontaine}
\affiliation{Centre for Quantum Technologies, National University of Singapore, Singapore 117543, Singapore}
\author{Adrian Copetudo}
\affiliation{Centre for Quantum Technologies, National University of Singapore, Singapore 117543, Singapore}
\author{Radim Filip}
\affiliation{Department of Optics, Palack\'{y} University, 17. listopadu 1192/12, 77146 Olomouc, Czech Republic}
\author{Yvonne Y. Gao}

\email[Corresponding author: ]{yvonne.gao@nus.edu.sg}
\affiliation{Centre for Quantum Technologies, National University of Singapore, Singapore 117543, Singapore}
\affiliation{Department of Physics,
National University of Singapore, Singapore 117542, Singapore}
\date{\today}

\begin{abstract} 
Quantum metrology offers the potential to surpass its classical counterpart, pushing the boundaries of measurement precision toward the ultimate Heisenberg limit. This enhanced precision is normally attained by utilizing large squeezed states or multi-particle entangled quantum states, both of which are often challenging to implement and prone to decoherence in real quantum devices.
In this work, we present a versatile and on-demand protocol for deterministic parameter estimation that leverages two efficient state-transfer operations on a single bosonic mode. Specifically, we demonstrate this protocol in the context of phase estimation using the superposition of coherent states in the bosonic circuit quantum electrodynamics (cQED) platform. 
With low average photon numbers of only up to 1.76, we achieve quantum-enhanced precision approaching the Heisenberg scaling, reaching a metrological gain of $7.5(6)$~dB. Importantly, we show that the gain or sensitivity range can be further enhanced on the fly by tailoring the input states, with different superposition weights, based on specific system constraints. The realization of this versatile and efficient scheme affords a promising path towards practical quantum-enhanced sensing, not only for bosonic cQED hardware but also readily extensible to other continuous-variable platforms.
\end{abstract}

\maketitle

\section{INTRODUCTION}
Quantum metrology utilizes 
optimally chosen sets of quantum states and measurement schemes to attain measurement precision beyond what is possible with classical states~\cite{giovannetti2004quantum}.
This quantum-enhanced precision appeals to many disciplines. 
For example, it has already applications for sensing magnetic fields~\cite{jones2009magnetic,napolitano2011interaction}, electric fields~\cite{facon2016sensitive,gilmore2021quantum}, and displacements~\cite{gilmore2021quantum,burd2019quantum}, as well as improving clock stability~\cite{ludlow2015optical,riedel2010atom} and quantum imaging~\cite{rouviere2024ultra, tsang2016quantum}. 
It is also expected to enhance gravitational wave detection in LIGO~\cite{tse2019quantum,yu2020quantum} and the search for dark matter~\cite{backes2021quantum}.  

One of the earliest and most canonical examples is quantum-enhanced phase estimation in a two-mode interferometer~\cite{nagata2007beating,slussarenko2017unconditional}. The use of coherent states (CS) defines the standard quantum limit (SQL) with precision $\Delta \theta \propto 1/\sqrt{N}$, where $\Delta \theta$ is the standard deviation of the phase and $N$ is the number of particles or excitations. Using entanglement as the resource state in this configuration, the precision can scale as $\Delta \theta \propto 1/N$ (known as the Heisenberg limit), which affords a $\sqrt{N}$ improvement over the SQL. This enhanced precision has been demonstrated in many experiments using entangled multipartite systems~\cite{guo2020distributed,li2023improving,marciniak2022optimal,mao2023quantum,franke2023quantum}. 
Practically, however, both the preparation of these non-local states and the implementation of correlated or conditional measurements can be challenging. Furthermore, preparing such quantum states often demands post-selection or heralding, which undermines the quantum enhancement~\cite{resch2007time,giovannetti2011advances} over deterministically created coherent states.

\begin{figure}[b!]
    \centering
    \includegraphics{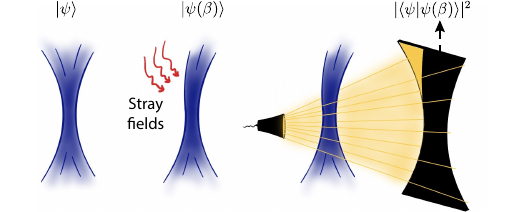}
    \caption{
    {\bf Generalized protocol for quantum metrology with a single bosonic mode.} Our approach consists of three fundamental steps: preparation of input state $|\psi\rangle$; distortion from the to-be-estimated process, e.g. stray electromagnetic fields, which effectively encode the parameter of interest $\beta$ into the state $|\psi(\beta)\rangle$; and final information retrieval by a projection operation that extracts the overlap with the initial state, shown as the shaded region. 
    }
    \label{fig1}
\end{figure}

In light of these challenges, there has been growing interest in quantum metrology using non-entangled states~\cite{braun2018quantum}. For instance, Fock states are used to perform phase-insensitive measurements with enhanced precision due to their nonclassical features~\cite{wolf2019motional,podhora2022quantum}.
Quantum-enhanced phase estimation has also been shown using displaced Fock states or their superposition. This handily bypasses the demanding non-local operations, which generally suppress the quantum advantage under even slight decoherence. However, existing single-mode schemes generally necessitate highly specialized state-dependent quantum gates, which are often expensive to implement~\cite{deng2023heisenberg,mccormick2019quantum,wang2019heisenberg} or require non-deterministic operations~\cite{deng2023heisenberg}.

\begin{figure*}[!th]
    \centering
    \includegraphics{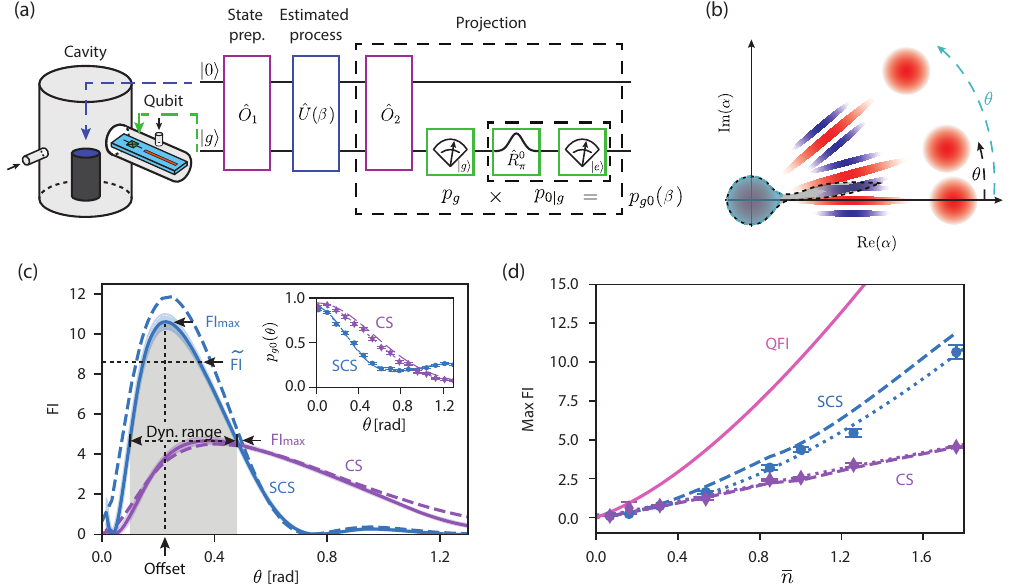}
    \caption{{\bf Parameter estimation procedure with bosonic cQED}.
    (a) Implementation of the protocol in cQED. A cavity stores a bosonic mode, coupled to a transmon qubit that facilitates operations and measurements. The readout resonator provides single-shot measurements of the qubit state. 
    The protocol includes a state-transfer operation $\hat O_1$, a process $\hat U(\beta)$, and a projection operation involving a state-transfer operation $\hat O_2$, qubit measurement $p_g$, and cavity measurement $p_{0|g}$.
    (b) Illustration of phase estimation using the SCS. The shaded area visualizes the overlap between the distorted and initial state $|\langle \psi|\psi(\theta)\rangle|^2$.
    (c) The FI for SCS (solid blue) and CS (solid purple), calculated using Eq.~(\ref{EQ_FI}), where $p$ is from the polynomial fit (dotted curves) of the experimental $p_{g0}(\theta)$ (markers) in the inset.
    The shaded area with the corresponding color indicates the standard deviation error from bootstrapping.
    Dynamical range is defined as the range on which the FI of SCS is above the maximum FI ($\rm{FI}_{max}$) of CS.
    The grey area represents the region where we have a quantum advantage compared to CS and $\tilde{\rm{FI}}$ is its corresponding average FI.
    Dashed curves show simulation.
    (d) The $\rm{FI}_{max}$ for SCS (blue) and CS (purple) are plotted as markers for states with different average photon numbers $\overline{n}$.
    The errors are standard deviation from bootstrapping.
    Dashed curves are simulation and dotted curves are fitting in the range $\bar n=[0.07,1.76]$: $3.92(9)\overline{n}^{1.71(1)}+0.15(4)$ for SCS and $2.62(6)\overline{n}^{0.99(1)}+0.00(4)$ for CS.
    The solid pink curve is the ideal QFI of SCS.}
    \label{fig2}
\end{figure*}

Here, we demonstrate a versatile and on-demand protocol for deterministic parameter estimation using a single mode with a standard bosonic cQED hardware. We show quantum-enhanced performance for phase as well as amplitude estimation using the superposition of coherent states (SCS), which minimizes the performance penalties due to hardware imperfections, achieving state-of-the-art metrological gain in the low photon number regime. We present a comprehensive analysis of key metrological figures of merit, including the dynamical range and demonstrate that our scheme can be adapted on the fly to optimize specific performance metrics. This is accomplished by varying a few programmable pulse parameters in the protocol to modify the superposition weights of the SCS. The efficiency and versatility of our implementation allow on-demand optimization of key performance metrics tailored to the metrological goal and specific hardware constraints. Finally, our strategy is adaptable across different physical systems and can be applied to other physical implementations of bosonic modes, such as optical resonators~\cite{haroche2006exploring}, trapped ions~\cite{leibfried2003quantum}, neutral atoms~\cite{grimm2000optical}, and mechanical resonators~\cite{aspelmeyer2014cavity}.

\begin{figure*}[!bt]
    \centering
    \includegraphics{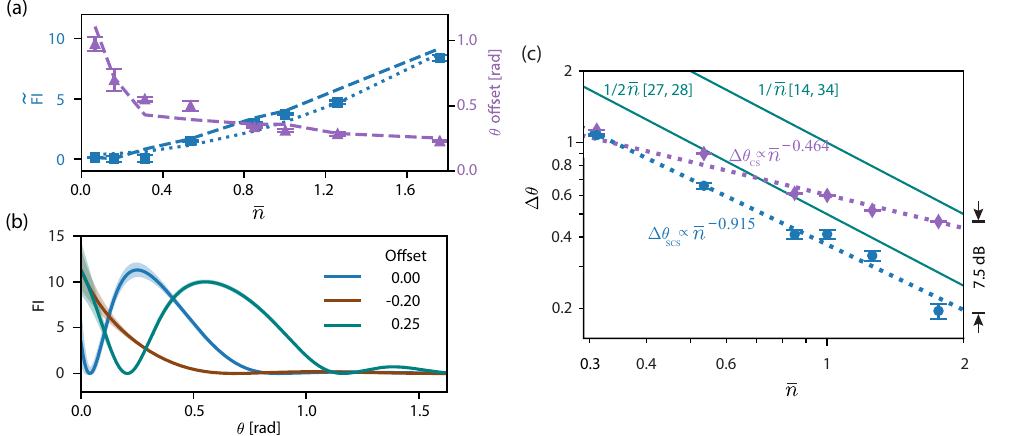}
    \caption{{\bf Comprehensive analysis for phase estimation and quantum-enhanced precision}. 
    (a) The average FI ($\tilde{\rm{FI}}$) of SCS within the dynamical range and $\theta$ offset at which FI is maximum vs $\overline{n}$. 
    The dotted curve is a fit: $2.7(3)\overline{n}^{2.0(1)}+0.3(1)$.
    (b) The same protocol as in Fig.~\ref{fig2}(c) with initial states prepared with offset in $\theta$.
    As the protocol is symmetrical with respect to $\theta$, maximum FI at $\theta=0$ can be obtained by a positive/negative offset of about $0.2-0.25$~rad.
    (c) The phase precision vs $\overline{n}$ for both SCS (blue circles) and CS (purple diamonds).
    The solid lines represent theoretical Heisenberg scaling with different factors.
    The dotted lines are linear fit in the range $\bar n=[0.07,1.76]$:  $\log(\Delta \theta)=-0.915(2)\log(\overline{n})-0.987(1)$ for SCS and $\log(\Delta \theta)=-0.464(1)\log(\overline{n})-0.508(1)$ for CS. The highest metrological gain for SCS compared to CS with the same photon number is $7.5(6)$~dB.
    }
    \label{fig3}
\end{figure*}

\section{THE PROTOCOL}
Conceptually, our methodology is akin to shining lights on an object under the influence of some external perturbations and comparing the resulting shadow with the original unperturbed object, as illustrated in Fig.~\ref{fig1}. The shadow represents the projection of the distorted state $|\psi(\beta)\rangle=\hat U(\beta)|\psi\rangle$ onto the initial state $|\psi\rangle$, written as $p(\beta)=|\langle \psi |\psi(\beta)\rangle|^2$ where $\beta$ is the to-be-estimated parameter, $\hat U(\beta)=\exp(i\hat B \beta)$, and we shall refer to $\hat B$ as the \emph{process generator}. The measurement outcome $p(\beta)$ provides an optimal precision for estimating $\beta$, see appendix~\ref{app:Optimality}, as indicated by the Fisher information (FI)
\begin{equation}\label{EQ_FI}
    F(\beta)=\frac{1}{p(\beta)(1-p(\beta))}\left(\frac{\partial p(\beta)}{\partial \beta}\right)^2,
\end{equation}
which provides a lower asymptotic bound on the achievable precision, i.e., $\Delta \beta \ge 1/\sqrt{F}$, known as the Cram\'{e}r-Rao bound~\cite{braunstein1994statistical}. The optimality transpires as the FI saturates its maximum value given by the quantum Fisher information (QFI) $F_q=4(\Delta \hat B)^2$, where $(\Delta \hat B)^2 = \langle \psi |\hat B^2 |\psi \rangle-\langle \psi |\hat B |\psi \rangle^2$ denotes the variance.

In addition to having an effective measurement scheme, it is also important to choose the input states strategically for optimal metrological performance. An immediate approach would be to consider states with high variance such as the superposition of Fock states~\cite{mccormick2019quantum,wang2019heisenberg}, the superposition of coherent states (SCS), and squeezed vacuum, see Appendix~\ref{app:Preliminary}. In practice, the choice of state must also be considered with the realistic hardware constraints in mind. In the bosonic cQED platform, the key sources of imperfection include the decoherence of the ancillary qubit, which reduces the quality of the prepared states, as well as the cavity self-Kerr effect, which distorts the state in phase space. The effects of both are more severe for states with higher Fock state occupations. This motivates the key advantage of our choice of using the SCS, which has higher variance compared to the superposition of Fock states and at the same time maintains a lower photon occupation in the cavity compared to squeezed vacuum, see Appendix~\ref{app:Preliminary}. Thus, our strategy effectively maximizes the relevant metrological figures of merit in the presence of hardware imperfections.

We implement this generalized metrology scheme on a standard cQED hardware as depicted in Fig.~\ref{fig2}(a). First, a desired quantum state is prepared with a state-transfer unitary $\hat O_1$, i.e., $|g\psi \rangle = \hat O_1|g0\rangle$, where $|g0\rangle=|g\rangle \otimes |0\rangle$ denotes the product state of the qubit-cavity system, with $|g\rangle$ the ground state of the qubit, and $|0\rangle$ the vacuum state of the cavity. After the system undergoes a process $\hat U(\beta)$,
a projection measurement is performed with a unitary $\hat O_2$ and qubit-cavity measurements. The operation $\hat O_2$ can be done with \emph{any} state-transfer unitary such that $|g0\rangle = \hat O_2|g\psi \rangle$. Finally, standard measurements are performed to obtain the probabilities of the qubit in the ground state $p_g$ and cavity in vacuum $p_{0|g}$, which is achieved by driving the qubit with a vacuum-selective $\pi$-pulse and measuring the qubit in the excited state. The projection yields the probability $p_{g0}=|\langle \psi |\hat U(\beta)|\psi \rangle|^2$, see Appendix~\ref{app:Optimality}.

An important advantage of our protocol is that any two operations $\hat O_1$ and $\hat O_2$ capable of realizing the required state transfers are sufficient. 
The second operation $\hat O_2$ has been implemented with time reversal technique~\cite{colombo2022time} ($\hat O_2=\hat O_1^{\dagger}$) or numerically optimized unitary~\cite{wang2019heisenberg}, both of which have to guarantee the evolution of multiple states concurrently. In contrast, our approach only requires constructing a single state-transfer process corresponding to $|g0\rangle = \hat O_2 |g\psi \rangle$. This provides an intuition on the appeal of our strategy, as there are infinitely many options for $\hat O_2$ that can perform the required state-transfer process whereas there is only one solution that satisfies $\hat O_2 = \hat O_1^{\dagger}$. As a result, finding an appropriate experimental operation, either numerically or analytically, to enact the general $\hat O_2$, instead of $\hat O_1^{\dagger}$, is significantly more efficient. 

Such a minimalist approach makes our protocol highly adaptable to a variety of hardware parameters and readily extensible to different physical systems with their preferred native gates. For instance, in cQED platforms featuring small qubit-oscillator dispersive coupling, $\hat O_{1,2}$ can be realized using echoed conditional displacement (ECD) gates~\cite{campagne2020quantum,eickbusch2022fast}; whereas devices with more significant dispersive coupling can leverage on gradient ascent pulse engineering (GRAPE)~\cite{ofek2016extending,heeres2017implementing,hu2019quantum} or selective number-dependent arbitrary phase gates~\cite{heeres2015cavity}. 
Beyond cQED, the state transfer operations can be achieved through coherent coupling of the internal-motional states of trapped ions/atoms~\cite{meekhof1996generation, mccormick2019quantum}, ECD gates in an atom-light system in cavity-QED~\cite{hacker2019deterministic}, or SWAP gates in qubit-mechanical resonator system in quantum acoustics~\cite{wollack2022quantum}.

\section{PHASE ESTIMATION}
We demonstrate the performance of this scheme in the context of phase estimation. The target process is enacted via the dispersive qubit-cavity coupling with Hamiltonian $H_{\text{int}}/\hbar = (\chi/2)\hat \sigma_z \hat a^{\dagger}\hat a$, where $\chi$ is the coupling strength, $\hat \sigma_z$ the Pauli-$z$ matrix, and $\hat a$($\hat a^{\dagger}$) is the annihilation (creation) operator of the bosonic cavity. For a qubit in the ground state $|g\rangle$, the dynamics realizes a unitary for the cavity $\hat U(\theta)=\exp(i\hat a^{\dagger}\hat a \theta)$, with $\theta = \chi t/2$. As a result, the process generator is given by $\hat n = \hat a^{\dagger}\hat a$, which means useful states here are those with high variance in photon number $(\Delta \hat n)^2$.

Our preliminary experimental investigations confirm that the SCS indeed offers the most robust sensitivity (highest FI) compared to the superposition of Fock states and squeezed vacuum, see Appendix~\ref{app:Preliminary}.
Therefore, we shall primarily focus on SCS, given by $\text{SCS}(\alpha)=\mathcal{N}_{\alpha}(|0\rangle +|\alpha \rangle )$, with the normalization factor $\mathcal{N}_{\alpha}$, to demonstrate the efficacy of our protocol. 
The SCS is simply a displaced cat state ($\mathcal{N}_{\alpha}(|-\alpha/2\rangle +|\alpha/2 \rangle )$), and in the limit of large $\alpha$ the normalization factor is $\mathcal{N}_{\alpha}\approx 1/\sqrt{2}$.
The intuition behind the performance of SCS for phase estimation is understood by visualizing the projection $p_{g0}=|\langle \psi |\hat U(\theta)|\psi \rangle|^2$ in phase space, see Fig.~\ref{fig2}(b).
With this illustration in mind, we note two important observations. Firstly, as $\theta$ increases, the probability $p_{g0}(\theta)$ changes from $1$ to a constant value, where the fringes do not overlap and the probability is insensitive to a further increase in $\theta$.
Secondly, the use of higher $\alpha$ (further right position of the right blob in Fig.~\ref{fig2}(b)) will reduce the overlap more sharply at lower $\theta$, making the detection more sensitive to $\theta$ at the cost of a smaller range of sensitivity.

Experimentally, we chose to implement the state transfer operations $\hat O_{1,2}$ using GRAPE, which is a convenient and widely used tool in cQED~\cite{ofek2016extending,heeres2017implementing,hu2019quantum,wang2019heisenberg}.
However, this choice is not unique, e.g., in the weak dispersive regime, one may opt to use ECD gates to create SCS~\cite{pan2023protecting}. By executing the gate sequence in Fig.~\ref{fig2}(a), we directly obtain the probability $p_{g0}(\theta)$. An exemplary plot comparing the use of SCS($\alpha=2$) and CS with the same average photon number $\overline{n}=\langle \hat a^{\dagger}\hat a \rangle$, is presented in Fig.~\ref{fig2}(c) inset. Each experimental point is an average of $1000$ single shot repetitions. The data shows good agreement with simulations (corresponding dashed curves), which take into account decoherences and measurement imperfections, see Appendix~\ref{app:Imperfections}. 

To quantitatively examine the performance of our implementation, we compute the FI in Eq.~(\ref{EQ_FI}), where the corresponding $p$ is a smooth function obtained from the polynomial fit of $p_{g0}(\theta)$. The resulting FI for SCS and CS are plotted in Fig.~\ref{fig2}(c) as solid blue and purple curves, respectively. As expected, the maximum FI of SCS beats that of the CS at the cost of the sensitivity range. Furthermore, figure~\ref{fig2}(d) shows the maximum FI ($\text{FI}_{\text{max}}$) for SCS (blue circles) and CS (purple diamonds) for states with different average photon numbers. We note that decoherence and experimental imperfections hinder our results from reaching the ideal QFI of SCS (pink solid curve), as shown by simulations (corresponding dashed curves) with real device parameters. However, despite that, second-order polynomial fits in the range $\bar n=[0.07,1.76]$ (corresponding dotted curves) show that the $\text{FI}_{\text{max}}$ of SCS is nearly quadratic, whereas that of the CS is linear, which indicates a clear $\sqrt{\overline{n}}$ quantum enhancement in the estimation precision. 

\section{METROLOGICAL FIGURES OF MERIT}
For a holistic evaluation of the performance of a quantum metrology scheme, several other important figures of merit must also be considered in conjunction with the maximum FI. Specifically, these include the parameter range in which we achieve quantum-enhanced precision, the corresponding average sensitivity, and the capability to shift the maximum sensitivity point to origin $\theta=0$ which is crucial for sensing applications. We define the dynamical range as the range of $\theta$ in which the FI of SCS is above the maximum FI of the CS, $\tilde{\text{FI}}$ as the average FI within the dynamical range, and $\theta$ offset where the FI is maximum. The dynamical range for SCS up to $\overline{n}=1.76$ is nearly constant at approximately 0.5~rad, which sets the limit on the allowed phase fluctuations in the system. The $\tilde{\text{FI}}$ against photon number $\overline{n}$ in Fig.~\ref{fig3}(a) shows a quadratic trend, suggesting that working within the dynamical range still affords favorable precision scaling. The offset in $\theta$ is close to the position of the maximum slope $|\partial p/\partial \theta|$ in the measured probability and decreases for SCS with larger $\alpha$, as shown in Fig.~\ref{fig3}(a). With our versatile protocol, this offset is a fully programmable parameter and can be adjusted by simply preparing input states with an initial phase offset (Fig.~\ref{fig3}(b)) to ensure that we can operate around the maximum sensitivity point.

In addition to the FI, which only sets the lower bound to the precision according to the Cram\'{e}r-Rao bound, we now present a benchmark for direct calculation of the phase precision, $\Delta \theta$. It is computed as $\Delta \theta = \Delta p/|\partial p/\partial \theta|$, where $\partial p/\partial \theta$ is derived from the polynomial fit as in Fig.~\ref{fig2}(c) inset and $\Delta p$ is the standard deviation of the single shot measurement with $1000$ repetitions. The standard deviation for SCS (blue circles) is plotted in Fig.~\ref{fig3}(c). The ideal precision, $\Delta \theta =1/\sqrt{4\overline{n}+4\overline{n}^2}$ (see Appendix~\ref{app:Preliminary}), surpasses the precision scaling $\Delta \theta=1/\overline{n}$ commonly used in optical~\cite{nagata2007beating,slussarenko2017unconditional} or atomic~\cite{colombo2022time} interferometers, as well as that achieved using superposition of Fock states ($\Delta \theta =1/2\overline{n}$)~\cite{mccormick2019quantum,wang2019heisenberg}. A linear fit of the results obtained using SCS also shows that $\Delta \theta_{\text{SCS}}\propto \bar{n}^{-0.915}$ in the range $\bar n=[0.07,1.76]$, which approaches the Heisenberg limit scaling $\Delta \theta \propto \bar{n}^{-1}$. To benchmark the quantum enhancement of our implementation, we compute the metrological gain of $20 \log_{10}(\Delta \theta_{\text{CS}}/\Delta \theta_{\text{SCS}})=7.5(6)$~dB with a low photon number of 1.76. Here, the uncertainty is one standard deviation obtained from bootstrapping and stabilizes at $\sim 600$ measurement repetitions, see Appendix~\ref{app:Amplitude}. With only minor adjustments to the protocol, we also showcase amplitude estimation with significant quantum enhancement, see Appendix~\ref{app:Amplitude}.

\begin{figure}[!t]
    \centering
    \includegraphics{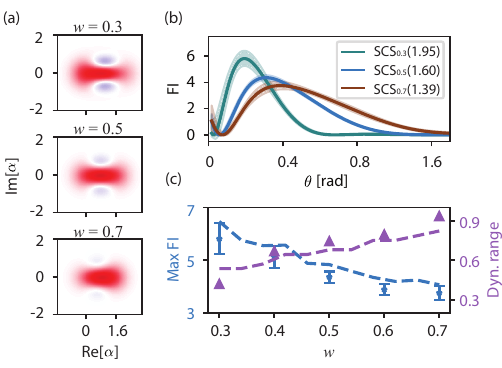}
    \caption{{\bf On-demand optimization for phase estimation}. (a) Wigner function plots of SCS$_w(\alpha)$ with different weights ($w$) but the same average photon number $\overline{n}=1$, 
    represented as $\mathcal{N}^{\prime}(\sqrt{1-w}|0\rangle+\sqrt{w}|\alpha\rangle)$, where $\mathcal{N}^{\prime}$ denotes the normalization coefficient.
    (b) FI versus phase for states with different weights $w$.
    The corresponding shaded area indicates the standard deviation from bootstrapping.
    (c) Maximal FI of SCS$_w(\alpha)$ and its dynamical range for different weights.  Dash curves: simulation results. Markers: experiment data.
    }
    \label{fig4}
\end{figure} 

\section{ON DEMAND OPTIMISATION SUBJECTED TO SYSTEM CONSTRAINTS}
Aside from the significant quantum enhancement achieved, our implementation also has the unique versatility of on-demand optimization for the specific hardware constraints present. Concretely, we show that the gain or dynamical range can be further enhanced without any hardware adjustment by optimizing the weights of the SCS. For instance, typical hardware implementations may often allow access to a limited maximum photon number $\overline{n}$ before the errors due to photon loss or the self-Kerr effect become too prominent. Physically, this is achieved by adjusting the qubit and cavity drives for $\hat O_{1,2}$, without any changes to the device or system dynamics. For the case where the constraint is the photon number, e.g., $\overline{n}=1$, we show that higher maximum FI can be achieved by putting more weight on the $|0\rangle$ component of the SCS, at the expense of sensitivity range, as shown in Fig.~\ref{fig4}(b) and (c).
Intuitively, if $\overline{n}$ is fixed, increasing the weight on the $|\alpha\rangle$ component comes at the cost of using a smaller $\alpha$, which leads to a reduction in the sensitivity. This trade-off between sensitivity and dynamical range as a function of the weights of the SCS is shown in Fig.~\ref{fig4}(c). This ability to perform such on-demand optimization is an intrinsic and powerful feature of our implementation, highlighting its practical feasibility for performing quantum metrology in the presence of realistic physical limitations in the hardware.

\section{CONCLUSIONS}
In summary, we have introduced a practical and versatile protocol for parameter estimation using a single bosonic mode. With this scheme, we demonstrated phase as well as amplitude estimation using the superposition of coherent states. Using states with average photon numbers of up to only 1.76, the high variance and robustness of SCS result in state-of-the-art metrological gain, $7.5(6)$~dB for phase and $9.3(5)$~dB for amplitude estimation, see Appendix~\ref{app:Amplitude}.
We also demonstrated the on-demand adaption of the scheme, utilizing SCS with different weights, to optimize for certain metrological performance metrics with respect to hardware constraints.  

The efficiency and versatility of our protocol stem from the use of a single bosonic mode with two state-transfer operations that can be realized by any standard gates. As such, our implementation is accessible to systems with different device parameters. Furthermore, the strategy is adaptable and can be readily performed with many other bosonic systems beyond cQED using their preferred native operations. Our results demonstrate a simple yet powerful strategy for real metrological applications. For instance, the phase estimation strategy can be adapted to directly estimate the coupling strength $\chi$ via Bayesian inference (see Appendix F). Future experiments may also extend this direction to estimate other quantities such as stray magnetic, electric field, etc. that may affect the qubit or cavity frequency, which in turn result in the change of dispersive coupling strength $\chi$~\cite{blais2021circuit}.
Therefore, our work offers a promising first step towards practical and optimal quantum metrology using a single bosonic system.

\section*{ACKNOWLEDGMENT}
This project is supported by the National Research Foundation, Singapore, grant number NRFF12-2020-0063 and NRF2020-NRF-ISF004-3540. R.F. acknowledges the support of the Czech Science Foundation via Grant number 21-13265X. K.P. acknowledges the project CZ.02.01.01/00/22$\_$008/0004649 (QUEENTEC) of EU and MEYS Czech Republic.

\appendix

\section{ Optimality of the protocol}
\label{app:Optimality}
The protocol consists of three parts: state preparation, estimated process and projection measurement as shown in the main text, in which state-transfer unitaries $\hat{O}_{1,2}$ can be done with any unitary for which the following holds: 
\begin{eqnarray}
|g\psi\rangle&=&\hat{O}_1|g0\rangle,\nonumber \\
|g0\rangle&=&\hat{O}_2|g\psi\rangle.
\end{eqnarray}
Note that the second line is equivalent to $\hat{O}_2^{\dagger}|g0\rangle=|g\psi\rangle$ and $\langle g0|\hat{O}_2=\langle g\psi|$.
Let us denote the final state before the qubit-cavity measurements as $|\psi_{\text{final}}\rangle=\hat{O}_2\hat{U}(\beta)\hat{O}_1|g0\rangle$ and the state with encoded parameter as $|g\psi(\beta)\rangle=\hat{U}(\beta)\hat{O}_1|g0\rangle$.
The probability after the qubit-cavity measurements, i.e., projection into $|g0\rangle$ can be expressed as
\begin{eqnarray}
p_{g0}&=&\langle g 0|\psi_{\text{final}}\rangle\langle\psi_{\text{final}}|g 0 \rangle \nonumber \\
&=&\langle g 0|\hat{O}_2\hat{U}(\beta)\hat{O}_1|g0\rangle\langle g0|\hat{O}_1^\dagger\hat{U}^\dagger(\beta)\hat{O}^\dagger_2 |g 0 \rangle \nonumber\\
&=&\langle g 0|\hat{O}_2|g\psi(\beta)\rangle\langle g\psi(\beta)|\hat{O}^\dagger_2 |g 0 \rangle \nonumber\\
&=&\langle g \psi|g\psi(\beta)\rangle\langle g\psi(\beta)|g \psi \rangle \nonumber\\
&=&|\langle \psi |\psi(\beta)\rangle |^2.
\end{eqnarray}
This overlap probability contains information regarding the parameter $\beta$ and is achieved with two state-transfer unitaries and standard qubit-cavity measurements.

To assess the sensitivity of the probability for parameter estimation, we compute the FI~\cite{braunstein1994statistical} for binary outcome:
\begin{equation}
    {\rm FI} = \frac{1}{p_{g0}(1-p_{g0})}\left(\frac{\partial p_{g0}}{\partial \beta}\right)^2.
    \label{eq_FI}
\end{equation}
The FI depends on particular measurement setting.
An optimal measurement setting allows the FI to reach the QFI $4(\Delta \hat B)^2$, which now only depends on the state and encoding operator of the corresponding parameter of interest.
We now show that our measurement setting is optimal, i.e., the FI saturates the QFI when $\beta\rightarrow 0$. 
In this case, we write $\hat{U}(\beta)=e^{i\hat{B}\beta}\approx \openone+i\hat{B}\beta-\hat{B}^2\beta^2/2-i\hat{B}^3\beta^3/6+\hat{B}^4\beta^4/24$ such that the probability reads
\begin{eqnarray}
    p_{g0} &=&|\langle\psi|\hat{U}(\beta)|\psi\rangle|^2 \nonumber \\
    &\approx &1-\beta^2(\Delta\hat{B})^2\nonumber \\
    &&+\beta^4\left(\frac{1}{12}\langle \hat B^4\rangle-\frac{1}{3}\langle \hat B\rangle\langle \hat B^3\rangle +\frac{1}{4}\langle \hat B^2\rangle^2\right).
\end{eqnarray}
Referring to Eq.~(\ref{eq_FI}), the FI can be written as 
\begin{eqnarray}
    \rm{FI}&=&\frac{1}{p_{g0}(1-p_{g0})}\left(\frac{\partial p_{g0}}{\partial\beta}\right)^2 \nonumber\\
    &\approx &4(\Delta\hat{B})^2 \nonumber \\
    &&-\beta^2\left(\langle \hat B^4\rangle-4\langle \hat B\rangle \langle \hat B^3\rangle+3\langle \hat B^2\rangle^2-4(\Delta \hat B)^2\right),\nonumber \\
\end{eqnarray}
where the absolute term independent of $\beta$ represents QFI. 

\begin{figure}{}
    \centering
    \includegraphics{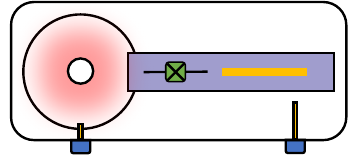}
    \caption{{\bf Schematic representation of a bosonic cQED hardware.} The device comprises three essential components: the cavity functions as the bosonic mode, the transmon introduces nonlinearity, and the resonator serves for single-shot readout. Additionally, one pin is coupled with the cavity for driving, while the transmon and resonator share the same channel for operation.}
    \label{fig:schematic}
\end{figure}

\section{device and system parameters}
\label{app:device}
Our device consists of a standard three-dimensional (3D) cavity, an ancillary transmon and a planar readout resonator, see Fig.~\ref{fig:schematic}. 
The cavity is made of high-purity (4N) aluminum and is a $\lambda/4$ cylinder-shaped transmission line surrounded by a wall. The readout resonator is deposited together with the transmon on the sapphire using a double-angle evaporation technique. The sapphire is inserted in the tunnel, with the pad slightly extending to the cavity. The cavity serves as a high-Q quantum memory to store a bosonic mode and is designed to dispersively couple with the ancillary transmon qubit, which provides the necessary  nonlinearity. The low-Q readout resonator is also dispersively coupled with the transmon qubit, enabling fast single-shot measurement. 

The Hamiltonian of the system can be written as:
\begin{equation}
    \hat{H}=\hat{H_0}+\hat{H_d},
\end{equation}
where $\hat H_0$ $(\hat H_d)$ incorporates the bare (drive) terms
\begin{equation*}
    H_0 = \Delta\hat{a}^\dagger\hat{a} -\chi \hat{a}^\dagger\hat{a}\hat{q}^\dagger\hat{q}- \frac{\alpha}{2} \hat{q}^\dagger\hat{q}^\dagger\hat{q}\hat{q}-
    \frac{K}{2}\hat{a}^\dagger\hat{a}^\dagger\hat{a}\hat{a}-\frac{\chi^{\prime}}{2} \hat{q}^\dagger\hat{q}\hat{a}^\dagger\hat{a}^\dagger\hat{a}\hat{a}
\end{equation*}
\begin{equation*}
    \hat{H}_d = i\zeta(t)\hat{a}^\dagger-i\zeta(t)^*\hat{a} + i\epsilon(t)\hat{q}^\dagger-i\epsilon(t)^*\hat{q}.
\end{equation*}

Here $\hat a$ ($\hat q$) denotes the annihilation operator of the cavity (transmon), $\Delta$ the frequency detunning between cavity and its drive, $\chi$ the dispersive coupling, $\chi^{\prime}$ the second order dispersive coupling, $\alpha$ the transmon anharmonicity, $K$ cavity anharmonicity or Kerr, $\zeta$ and $\epsilon$ respectively the drive strength of the cavity and transmon. 

The system parameters are summarized in Table~\ref{table:1}.

\begin{table}[h]
    \centering
    \begin{tabular}{|p{6.4cm}|p{0.9cm}|p{0.9cm}|}
        \hline
        Cavity frequency (\rm{GHz}) & $\omega_c/2\pi$ & 4.587 \\
        Qubit frequency (\rm{GHz}) & $\omega_q/2\pi$ & 5.277\\
        Resonator frequency (\rm{GHz}) & $\omega_r/2\pi$ & 7.617 \\
        Cavity single-photon lifetime (\rm{ms}) &$T_{c,1}$ & 1 \\
        Transmon energy lifetime $(\mu s)$ & $T_{q,1}$ & 96 \\
        Transmon coherence lifetime $(\mu s)$& $T_{q,2}$ & $15.47$ \\
        Cavity anharmonicity (\rm{kHz})  & $K/2\pi$ & $6$ \\
        Transmon anharmonicity (\rm{MHz})  & $\alpha/2\pi$ & $175.3$ \\
        1st cavity-transmon dispersive shift (\rm{MHz}) & $\chi/2\pi$ & 1.423 \\
        2nd cavity-transmon dispersive shift (\rm{kHz}) & $\chi^{\prime}/2\pi$ & 15.76\\
        Transmon-resonator dispersive shift (\rm{MHz}) & $\chi_{qr}/2\pi$ & 0.64 \\
        \hline
    \end{tabular}
    \caption{{\bf System parameters.} Summary of the key system parameters and lifetimes.}
    \label{table:1}
\end{table}

\begin{figure}
    \centering
    \includegraphics{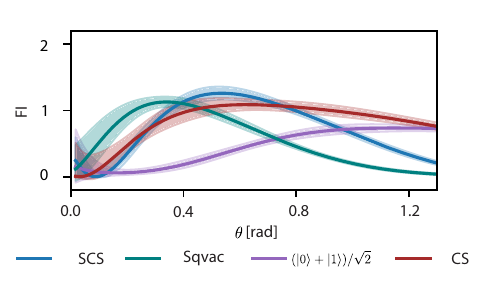}
    \caption{{\bf Phase estimation utilizing various states.}
    The FI of phase estimation  utilizing  CS, superposition of Fock states, squeezed vacuum (Sqvac) and SCS with fixed average photon number $\bar n=0.5$.}
    \label{fig:diffstates}
\end{figure}

Note that for sufficiently high anharmonicity, the transmon can be effectively treated as a qubit (two-level system with $|g\rangle$ ($|e\rangle$) as the ground (excited) level).
The first two terms in the bare Hamiltonian then becomes $\Delta\hat{a}^\dagger\hat{a} -\chi \hat{a}^\dagger\hat{a}|e\rangle \langle e|$, which reduces to $ (\chi/2)\hat \sigma_z \hat a^{\dagger}\hat a$ for $\Delta=\chi/2$ that is chosen.

\section{ Preliminary investigations using different states}
\label{app:Preliminary}
Useful states for achieving low estimation error $\Delta \beta$ encoded with an operator $\hat B$ via $\hat U(\beta)=e^{i\hat B\beta}$ are those with high variance $(\Delta \hat B)^2$.

In the context of phase estimation, we investigate several states such as CS $|\alpha \rangle$, superposition of Fock states $(|0\rangle +|N\rangle)/\sqrt{2}$, squeezed vacuum $e^{\frac{-r}{2}(\hat a \hat a- \hat a^{\dagger}\hat a^{\dagger})}|0\rangle$, and SCS $\mathcal{N}_\alpha(|0\rangle +|\alpha \rangle)$.
Here, the normalisation factor $\mathcal{N}_\alpha=1/\sqrt{2(1+e^{-|\alpha|^2/2})}$.
The average photon number $\overline{ n}=\langle \hat a^{\dagger}\hat a\rangle$ of these states are respectively $|\alpha|^2$, $N/2$, $\sinh^2{r}$, and $\mathcal{N}_\alpha^2|\alpha|^2$.
The QFI $4(\Delta \hat n)^2$ of these states in terms of $\bar n$ are respectively $4\overline{n}$, $4\overline{n}^2$, $8\overline{n}^2+8\overline{n}$, and $4((1-\mathcal{N}_\alpha^2)/\mathcal{N}_\alpha^2)\overline{n}^2+4\overline{n}$.
The QFI of SCS reduces to $4\overline{n}^2+4\overline{n}$ for large $\overline{n}$ as $\mathcal{N}_\alpha\approx 1/\sqrt{2}$.

Experimentally, we realize these states and compute the corresponding FI with the same method as described in the main text, using Eq.~(\ref{eq_FI}), see Fig.~\ref{fig:diffstates}.
The comparison is made for states with the same photon number $\overline{n}=0.5$. 
We also computed the Fisher information using four probabilities by considering the qubit measurement outcome in both ground and excited states: $p_{g0}(\theta), 1-p_{g0}(\theta), p_{e0}(\theta), 1-p_{e0}(\theta)$. This way, the Fisher is computed as $\mbox{FI}=\sum_i (1/p_i)(\partial p_i/\partial \theta)^2$, where $\{p_i\}$ are the four probabilities listed above.
This method can enhance the FI only by a factor of $\approx 1.2$ compared to using Eq.~(\ref{eq_FI}), despite requiring twice the number of measurements.
Therefore, for the rest of the analysis, we will compute Fisher information using Eq.~(\ref{eq_FI}).

It is apparent from Fig.~\ref{fig:diffstates} that the SCS offers the highest FI.
Although the squeezed vacuum is supposed to give a higher theoretical QFI, the state is harder to generate as it requires high squeezing.
Highly squeezed states exhibit increased spreading in phase space, making them susceptible to the Kerr effect, which, in turn, hinders quantum-enhanced sensitivity. A viable strategy to mitigate the impact of the Kerr effect involves minimizing the coupling between the cavity and transmon. By reducing this coupling, the undesirable consequences of the Kerr effect can be alleviated, contributing to enhanced quantum sensitivity.
Therefore, we discern that with our device, the optimal states are SCS.

\begin{figure}
    \centering
    \includegraphics{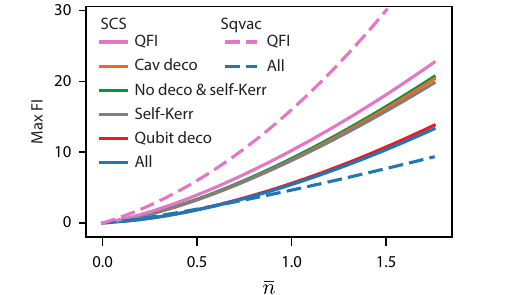}
    \caption{{\bf Effect of imperfections on the maximum Fisher information obtained using SCS and squeezed vacuum state via full Hamiltonian simulation.}
    The maximum Fisher information from simulations as a function of the average photon number $\bar{n}$ for SCS (solid curves), considering individual imperfections independently. The pink solid curve is the theoretical QFI. For comparison, we also show the results obtained using squeezed vacuum states (dashed curves).}
    \label{fig:errorbudget}
\end{figure}

\begin{figure*}
    \centering
    \includegraphics{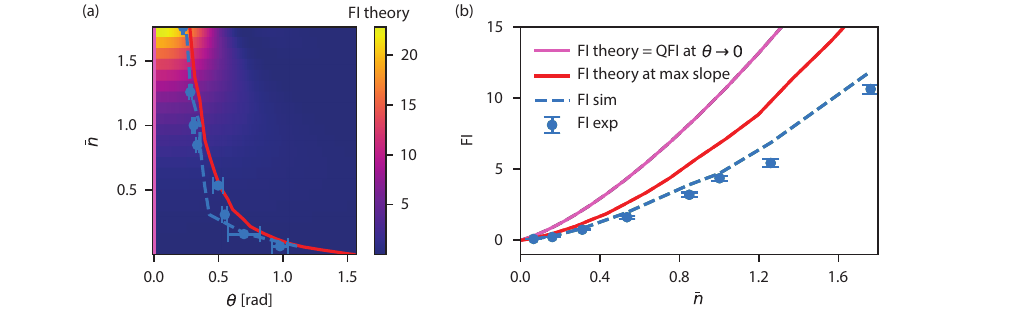}
    \caption{{\bf Position of maximum sensitivity.}
    (a) Theoretical FI for SCS vs $\theta$ for states with different average photon number $\bar n$.
    The pink and red lines are FI at $\beta \rightarrow 0$ and maximum probability slope $|\partial p/\partial \theta|$ respectively.
    Experimental data (blue circles) and simulation (blue dash curve) from Fig.~3(a) in the main text are replotted here  for comparison.
    (b) The corresponding different FI are plotted for comparison.}
    \label{fig:thetaoff}
\end{figure*}

\section{Imperfections in the system}
\label{app:Imperfections}

We use gradient ascent pulse engineering (GRAPE) pulse to realize the two state-transfer operations in the protocol. The GRAPE pulses ($\zeta(t)$ and $\epsilon(t)$) are optimized in an ideal system without imperfections.
In practice, each element of the system is susceptible to environmental noise, making it crucial to account for the coherence and energy decay rate of both the cavity and transmon qubit.
Table~\ref{table:1} presents the corresponding parameters. 
To account for decoherence, the simulations in the main text (all the sequence) are performed with the following master equation:

\begin{equation}
    \frac{d \rho}{dt}=-\frac{i}{\hbar}[\hat{H}, \rho]+\sum_k \hat L_k\rho\hat L_k^\dagger-\frac{1}{2}(\hat L_k^\dagger\hat L_k\rho+\rho\hat L_k^\dagger\hat L_k),
\end{equation}
where $\hat L_k$ denotes the jump operator for the corresponding decoherence.
We consider five types of decoherence: qubit relaxation $\hat L_1 = \hat{q}\sqrt{(1+ n_q)/T_{q,1}}$,  qubit thermal excitation $\hat L_2 = \hat{q}^\dagger \sqrt{n_q/T_{q,1}}$, qubit dephasing $\hat L_3 = \hat{q}^\dagger\hat{q}\sqrt{2/T_{q,\phi}}$, cavity relaxation $\hat L_4 = \hat{a}\sqrt{(1+n_c)/T_{c,1}}$, cavity thermal excitation $\hat L_5 = \hat{a}^\dagger \sqrt{n_c/T_{c,1}}$, where $T_{q,\phi}=1/(1/T_{q,2}-0.5/T_{q,1})$ and $n_c$ ($n_q$) is the average thermal excitations in the cavity (qubit). 

The dashed curves in Fig.~2(d) represent simulations that account for all imperfections. To investigate the individual impact of each imperfection on the maximum FI, we perform simulations in which each imperfection is considered independently. The solid curves in Fig.~7 illustrate the theoretical QFI (pink) and the maximum FI as a function of the average photon number $\bar{n}$ for several cases using SCS: green (no decoherence and no self-Kerr), orange (cavity $T_{c,1}$), gray (self-Kerr), red (qubit $T_{q,1}$ and $T_{q,2}$), and blue (all imperfections).
Notably, the green curve does not closely saturate the QFI. This can be attributed to imperfect state preparation via GRAPE pulses, which achieves a fidelity of approximately $0.998$. As expected, simulations using ideal initial SCS in place of the one generated via GRAPE pulse (not shown) confirms that the outcome saturates the QFI. 
Cavity decay (solid orange curve) results in a minor reduction in the FI due to the cavity's long lifetime ($\approx 1$ ms). The self-Kerr effect, an unwanted term in the Hamiltonian that distorts the state in phase space during the phase-encoding process, causes a slightly larger decrease in the FI (solid gray curve). The most significant degradation arises from the qubit decoherence (solid red curve). During the dynamics, numerical pulses entangle the qubit and cavity states to facilitate operations on the cavity state, but at the same time making it vulnerable to qubit energy decay and dephasing. Finally, the solid blue curve incorporates all imperfections.
For comparison, we also include results for squeezed vacuum states: dashed pink (QFI) and dashed blue (all imperfections). Although squeezed vacuum states theoretically provide higher QFI, they are more susceptible to decoherence experimentally, with the FI (dashed blue curve) losing its quadratic dependence and appearing almost linear under the realistic hardware imperfections considered here.

\begin{figure*}
    \centering
    \includegraphics{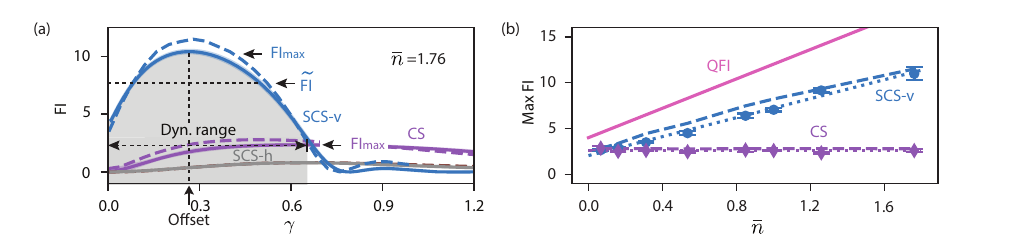}
    \caption{{\bf Amplitude estimation}.
    (a) FI for amplitude estimation using CS (solid purple), SCS-v for amplitude in the vertical direction (solid blue), and SCS-h for amplitude in the horizontal direction (solid grey).
    The corresponding dashed curves indicate simulation.
    (b) The maximum FI for SCS-v (blue) and CS (purple) vs photon number $\overline{n}$.
    Dashed curves are simulation and dotted curves are fitting: $4.9(2)\overline{n}^{1.06(2)}+2.2(1)$ for SCS-v and $0.411(1)\overline{n}^{0.00(3)}$ + 2.165(1) for CS.
    The solid pink curve is the ideal QFI of SCS-v. }
    \label{fig:amplitide}
\end{figure*}

\begin{figure*}
    \centering
    \includegraphics{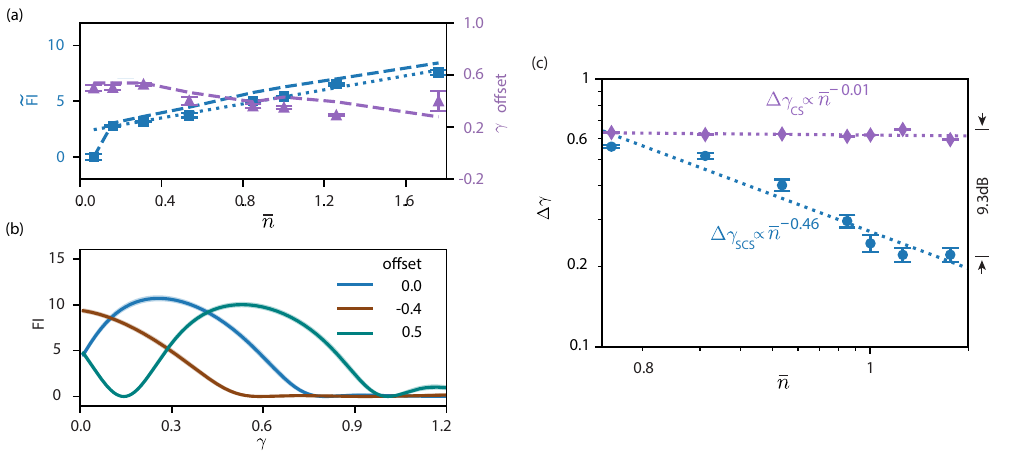}
    \caption{{\bf Performance of quantum-enhanced amplitude precision}. 
    (a) The average FI of SCS for which it is above the max FI of CS and $\gamma$ offset at which FI is maximum vs $\bar n$. The dotted line is linear fit of $3.2(2)\overline{n}^{1.0(3)}+2.2(1)$.
    (b) The same protocol as in Fig.~9(a) with initial states prepared with an offset.
    (c) The amplitude precision vs $\bar n$ for both SCS (blue circles) and CS (purple diamonds).
    The dotted lines are linear fit of $\log(\Delta \gamma)=a\log(\overline{n})+b$ with $a$ and $b$ as free parameters.
    The fitted parameters are $a,b=-0.460(3), -1.311(2)$ for SCS and $a, b=-0.0105(2), -0.480(1)$ for CS. }
    \label{fig:comp_amplitude}
\end{figure*}

In Appendix A, we see that ideally maximum FI is achieved at $\beta \rightarrow 0$.
In the presence of imperfections, this is not the case.
To exemplify this let us first compute the FI of SCS for phase estimation.
The probability of overlap is given by
\begin{eqnarray}
    p &=& |\langle \psi|\psi(\theta)\rangle|^2\nonumber \\
    &=&\frac{1}{4(1+e^{-\frac{\alpha^2}{2}})^2}|1+2e^{-\frac{\alpha^2}{2}}+e^{-\alpha^2(1-e^{i\theta})}|^2.\label{EQ_overlapth}
\end{eqnarray}
From $p$, the theoretical FI can be derived analytically.
This FI is plotted in Fig.~\ref{fig:thetaoff}(a) vs $\theta$ for SCS with different average photon number $\bar n$.
We note that at $\beta \rightarrow 0$ (pink line), the FI saturates the QFI.
We also plotted the FI at the maximum slope $|\partial p/\partial \theta |$ as red line.
We note that the experimental data (blue circles) follows this path closely.
This supports a strategy that the practical position of maximum sensitivity is around the maximum slope of probability.
The comparison of the corresponding FI is plotted in Fig.~\ref{fig:thetaoff}(b).

\section{ Amplitude estimation using the superposition of coherent states}
\label{app:Amplitude}

To illustrate the versatility of our protocol, we demonstrate quantum-enhanced precision for amplitude estimation by simply changing the process unitary in Fig. 2(a) in the main text to a displacement in the imaginary direction $\hat U(\gamma)=\exp(i\gamma(\hat a^{\dagger}+\hat a))$, where $\gamma$ is the displacement amplitude. The process generator is now $\hat x = (\hat a^{\dagger}+\hat a)$, making states with high variance $(\Delta \hat x)^2$ useful for amplitude estimation. Here, SCS offer high variance $(\Delta \hat x)^2 = 2\overline{n} +1$ compared to CS, which have a constant variance $(\Delta \hat x)^2 = 1$. The illustration is similar to that of phase estimation of  Fig. 2(b) in the main text, except the SCS are now displaced in the imaginary direction instead of rotated. SCS with higher $\alpha$ have more rapidly oscillating fringes such that the overlap with the initial state reduces more sharply as $\gamma$ increases.

\begin{figure}
    \centering
    \includegraphics{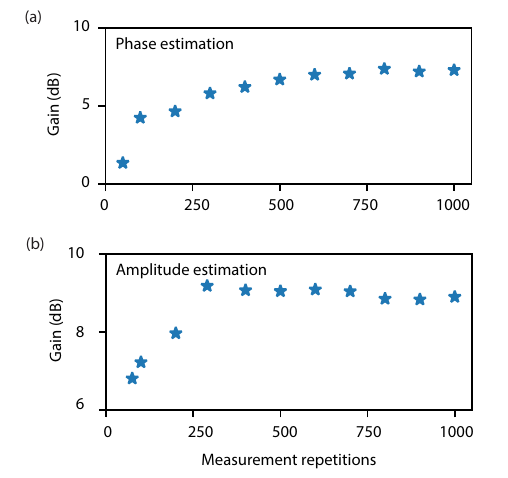}
    \caption{{\bf Metrological gain vs measurement repetitions.} The precision gain of SCS over CS for (a) phase and (b) amplitude estimation for different numbers of measurement repetitions.}
    \label{sfig5}
\end{figure}

Experimentally, we show that SCS displaced vertically in the imaginary direction (SCS-v) offer the highest FI, compared to both CS and SCS displaced horizontally in the real direction (SCS-h), as shown in Fig.~\ref{fig:amplitide}(a). This can be explained by noting that SCS-h has a different process unitary $\hat U(\gamma)=\exp(i\gamma i(\hat a-\hat a^{\dagger}))$, where the process generator is now $\hat y=i(\hat a-\hat a^{\dagger})$, such that the variance of SCS is low, even lower than that of the CS. 

Furthermore, by operating at the maximum FI, we show that the SCS-v (CS) exhibits a linear (constant) trend against $\overline{n}$, as shown in Fig.~\ref{fig:amplitide}(b), indicating $\sqrt{\overline{n}}$ quantum enhancement in precision.

Similar to phase estimation, it is essential to gain insights into the mean sensitivity above maximum FI of CS and the $\gamma$ offset at which the FI is optimal. 
With the increase of photon number, the average FI increases linearly while the optimal amplitude sensor point is moving to 0 as shown in Fig.~\ref{fig:comp_amplitude}(a). 
The optimal amplitude sensor point can also be shifted to 0 by adding a displacement to the initial state in the GRAPE pulse, which is crucial for direct amplitude sensing. 
Furthermore, to estimate the amplitude estimation error directly via $\Delta \gamma = \Delta p/|\partial p/\partial \gamma|$, we present a benchmark in Fig.~\ref{fig:comp_amplitude}(c). 
The amplitude estimation error for the CS remains constant even with an increase in the average photon number. In contrast, the estimation error using SCS exhibits an almost linear relationship with the average photon number and shows a maximum gain of 9.3(5)dB.

Finally, an important factor to consider for sensing applications is the measurement repetitions, which also directly translates to the total duration of the protocol.  
The highest metrological gain for phase and amplitude estimation we report in the main text was done with $1000$ measurement repetitions.
The respective gain vs measurement repetitions are plotted in Fig.~\ref{sfig5}.
Notably, the gain for both phase and amplitude estimation stabilizes at $\sim 600$ measurement repetitions.
It shows that beyond this point, the single-shot measurements are enough to provide meaningful statistics.

\section{Bayesian inference}
\label{app:Bayesian}

Bayesian estimation is a powerful statistical tool that updates the probability estimate for a hypothesis as more evidence or information becomes available~\cite{Paris2004}. In the context of quantum metrology, Bayesian inference allows for the optimal estimation of parameters by systematically incorporating experimental data to refine prior knowledge. Here we first briefly recap the Bayesian method and then utilize it to perform phase estimation $\theta$ with our experimental data using SCS to recover the results and trends presented in the main text. In addition, we also perform Bayesian inference to directly estimate the dispersive coupling strength $\chi$.

\subsection{The Bayesian framework}

The Bayesian approach hinges on Bayes' theorem, which relates the posterior probability $P(\theta|\mathcal{D})$ of a parameter $\theta$ given data $\mathcal{D}$ to the prior probability $P(\theta)$ and the likelihood $P(\mathcal{D}|\theta)$:
\begin{equation}
    P(\theta|\mathcal{D}) = \frac{P(\mathcal{D}|\theta) P(\theta)}{P(\mathcal{D})},
\end{equation}
where $P(\mathcal{D})$ is the evidence, ensuring the posterior is properly normalized.

In our study, the parameter of interest is the phase $\theta$, and the data $\mathcal{D}$ comprises the measurement outcomes from experiments using SCS probes.

For each experimental trial, the overlap $p_{g0}(\theta)=|\langle \psi |\psi(\beta)\rangle|^2$ is measured in a single-shot manner, yielding outcomes $1$ or $0$ whose probabilities shall be respectively denoted henceforth as $p_1(\theta)=p(\theta)$ and $p_0(\theta) = 1 - p(\theta)$ for ease of notation. The likelihood function for observing $N_1$ number of outcome $1$ in $N$ trials is given by:
\begin{equation}
    L(\theta) = p_1(\theta)^{N_1} \, (1-p_1(\theta))^{N - N_1}.\label{EQ_likelihood}
\end{equation}

Assuming a uniform prior $P(\theta)$ over the interval $\theta \in [0, 2.5]$, the posterior distribution simplifies to being proportional to the likelihood:
\begin{equation}
    P(\theta|\mathcal{D}) \propto L(\theta).\label{EQ_posterior}
\end{equation}

The Bayesian estimate, $\theta_{\text{Bayesian}}$, is obtained by maximizing the posterior distribution, and the uncertainty $\Delta \theta$ is given by the standard deviation of the posterior.

\begin{figure*}[htbp]
    \centering
    \includegraphics[width=1.0\linewidth]{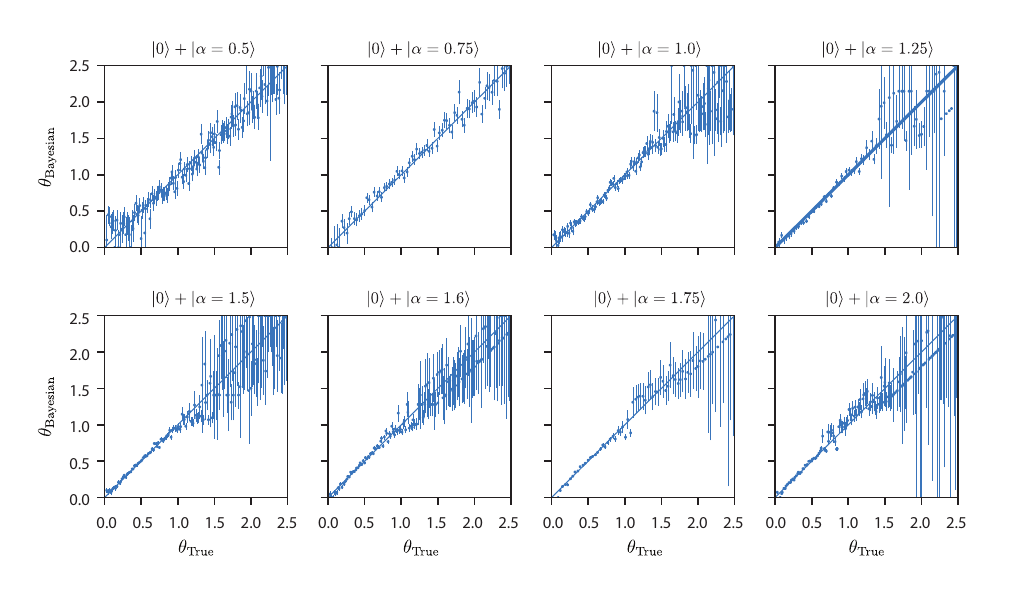}
    \caption{
        Comparison of the estimated phase via Bayesian inference $\theta_{\text{Bayesian}}$ with the true phase $\theta_{\text{True}}$ for various SCS amplitudes $\alpha$. Error bars represent the standard deviation of the posterior distribution. The solid lines correspond to equality. }
    \label{fig:figtotcorr}
\end{figure*}

\begin{figure*}[htbp]
    \centering
    \includegraphics[width=1.0\linewidth]{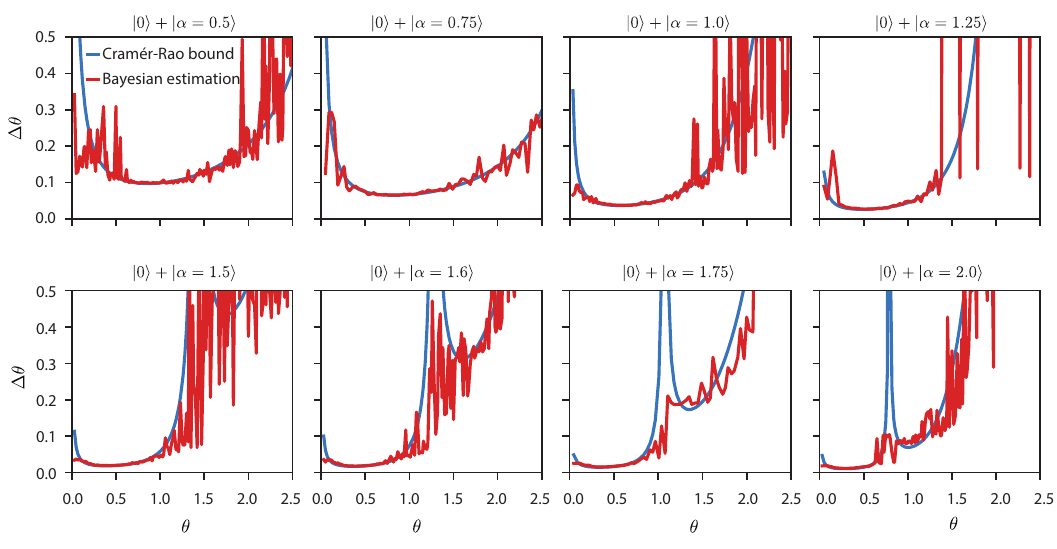}
    \caption{
        Uncertainty in the estimation $\Delta \theta$ as predicted by the Cram\'{e}r-Rao bound (blue) and Bayesian inference (red). Different panels correspond to various coherent amplitudes $\alpha$ of the SCS. The Bayesian estimates generally align with the Cram\'{e}r-Rao bound predictions within the dynamical range, validating the efficiency of the Bayesian approach in approaching the Cram\'{e}r-Rao bound.}
    \label{fig:estimation}
\end{figure*}

\subsection{Phase estimation}

In order to get the probability $p_1(\theta)$, we fit the experimental data with the probability model inspired by the theoretical overlap in Eq.~(\ref{EQ_overlapth}):
\begin{align}
p_1(\theta)=\frac{\left(c+e^{-{\alpha'} ^ 2 \left(1-e^{-i \theta }\right)}\right) \left(c+e^{-{\alpha'}^2 \left(1-e^{i \theta }\right)}\right)}{4 b},\label{EQ_fitprob}
\end{align}
where the free parameters $b$, $c$, and $\alpha'$ (effective amplitude of the coherent state) account for experimental imperfections. This fitted probability function, together with the single-shot experimental data $N_1$ at each phase $\theta$ allows the calculation of $\theta_{\text{Bayesian}}$ and the standard deviation $\Delta \theta$ from the posterior distribution given in Eq.~(\ref{EQ_posterior}).

We illustrate the alignment between the Bayesian estimates and the true phase across different $\alpha$ values in Fig.~\ref{fig:figtotcorr}. As $\alpha$ increases, the precision of the estimation (within the dynamical range) improves, which is evident from the reduced error bars. However, this comes with a trade-off in the dynamical range, limiting the range of $\theta$ over which the estimation remains accurate.

\begin{figure}[htbp]
    \centering
    \includegraphics[width=1.0\linewidth]{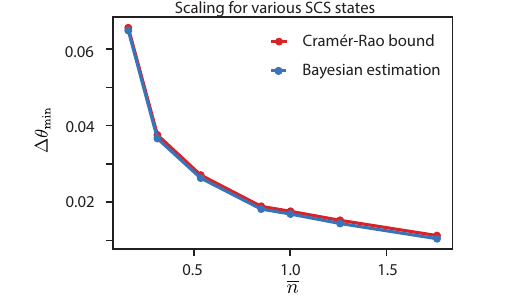}
    \caption{
        Minimal estimation uncertainty $\Delta \theta_{\text{min}}$ plotted against the average photon number $\bar n$ for various SCS. Both Bayesian estimation and Cramér-Rao bound predictions show a similar decreasing trend, confirming the scaling relationship between precision and $\bar n$.}
    \label{fig:scaling}
\end{figure}

Next, we compare the uncertainty $\Delta \theta$ derived from Bayesian estimation with the Cramér-Rao bound prediction via $\Delta \theta = 1/\sqrt{N \, \text{FI}}$ in Fig.~\ref{fig:estimation}. The close agreement between the two methods, within the dynamical range, demonstrates the effectiveness of Bayesian inference in achieving near-optimal parameter estimation. The minimal uncertainty $\Delta \theta_{\text{min}}$ as a function of the coherent state amplitude $\alpha$ (or average photon number $\bar n$) exhibits a decreasing trend, as shown in Fig.~\ref{fig:scaling}, indicating enhanced precision with larger $\bar n$. This scaling is consistent with theoretical predictions based on the Cram\'{e}r-Rao bound.

\subsection{Estimation of $\chi$}

The phase $ \theta $ is linearly related to the interaction strength $ \chi $ by $ \theta = \chi t/2 $, assuming the time $ t $ is precisely known and controlled. In this framework, the uncertainty in $ \theta $ directly translates to the uncertainty $ \Delta \chi $, and selecting the optimal $ t $ corresponds to optimizing $ \theta $ for sensitivity.

An iterative strategy can be used to refine the estimate of $ \chi $. Initially, an arbitrary choice of $ t^{(1)} $ is used for the first measurement round. From the resulting experimental data, a Bayesian estimate $ \chi^{(1)} $ is obtained. In the subsequent iteration, the interaction time is adjusted to $ t^{(2)} $, chosen such that $ \chi^{(1)} t^{(2)}/2 \approx \theta_\mathrm{opt} $, where $ \theta_\mathrm{opt}$ is the point of maximum sensitivity (at which $\text{FI}(\theta)$ is maximum). This process repeats, with each iteration refining the estimate $ \chi^{(k)} $ and updating the interaction time $ t^{(k+1)} $ accordingly. The iterations continue until the desired precision in $ \chi $ is achieved, leveraging Bayesian inference at each step to optimize and update the experimental parameters dynamically.

For the subsequent iteration, we can also make use of all the data accumulated in the previous rounds. As such, we estimate $\chi$ with Bayesian method utilizing the data of different $t$'s.
Let us consider $N_t$ discrete times $\{ t_i \}_{i=1}^{N_t}$. At each time $t_i$, we perform $N_i$ measurements and record the number of times the measurement yields outcome $1$ as $N_1[t_i]$, with the number of $0$ detections being $N_0[t_i] = N_i - N_1[t_i]$.
The probability of detecting the outcome $1$ at time $t_i$ given the interaction strength $\chi$ is
\begin{equation}
    p_1(t_i | \chi) = p_1\left( \theta_i \right) = p_1\left( \frac{\chi t_i}{2} \right),
    \label{eq:prob_detection}
\end{equation}
where $p_1(\theta)$ is the probability function derived via Eq.~(\ref{EQ_fitprob}), as discussed earlier. The likelihood function for $\chi$ based on all the measurement outcomes is then constructed as
\begin{equation}
    \mathcal{L}(\chi) = \prod_{i=1}^{N_t} \left[ p_1(t_i | \chi) \right]^{N_1[t_i]} \left[ 1 - p_1(t_i | \chi) \right]^{1-N_1[t_i]}.
    \label{eq:likelihood_chi}
\end{equation}

Assuming a uniform prior for $\chi$ over a plausible range, the posterior probability distribution is proportional to the likelihood function:
\begin{equation}
    \mathcal{P}(\chi) \propto \mathcal{L}(\chi).
    \label{eq:posterior_chi}
\end{equation}
The maximum of $\mathcal{P}(\chi)$ yields the most probable estimate $\chi_{\text{est}}$, and the standard deviation of the posterior distribution provides the uncertainty in the estimation $\Delta \chi$.

\begin{figure}[htbp]
    \centering
    \includegraphics[width=\linewidth]{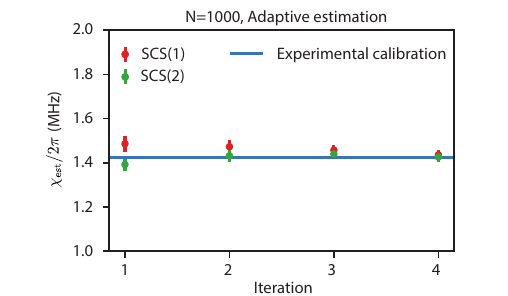}
    \caption{
                Convergence of the estimated interaction strength $ \chi_{\text{est}} $ as a function of iteration steps in the adaptive estimation procedure. The probe states are the SCS with $\alpha = 1$ and $2$, with $N = 1000$ new sampled data per iteration. The rapid reduction in estimation error illustrates the efficiency of the adaptive Bayesian approach. The higher value of $ \alpha $ further enhances the estimation precision, leading to faster convergence to the true value.}
    \label{fig:chivsalpha}
\end{figure}

We illustrate the adaptive estimation process for the interaction strength $ \chi $ over multiple iteration steps in Fig.~\ref{fig:chivsalpha}. The estimated value rapidly converges to the independently calibrated value of $\chi$, shown as the blue line. At each iteration, the estimation leverages all accumulated data, leading to progressively improved precision. The figure highlights the robustness and efficiency of the adaptive Bayesian approach in achieving accurate and reliable estimates of $ \chi $. For example, with only 4 iterations, $\Delta \chi/2 \pi \approx 0.024$ MHz ($1.7\%$) for SCS with $\alpha=2$. Better precision can be obtained by using more iterations or using SCS with higher amplitude $\alpha$.

\newpage

\bibliography{Bibliography}   

\end{document}